\documentclass[pra,aps,twocolumn,superscriptaddress,nofootinbib,nopacs]{revtex4} 

\usepackage{amsmath}  \usepackage{amssymb}  \usepackage{amsfonts}  \usepackage{bm}  \usepackage{bbm}   \usepackage{braket}  \usepackage{color}  \usepackage{comment}  \usepackage{dcolumn}  \usepackage{enumerate}  \usepackage{epsfig}  \usepackage{gensymb}  \usepackage{graphicx}  \usepackage{indentfirst}  \usepackage{lmodern}  \usepackage{mathrsfs}  \usepackage{mathtools}  \usepackage{psfrag}  \usepackage{pst-all}  \usepackage{soul}  \usepackage{units}  \usepackage{xcolor}
\usepackage{float} 
\usepackage[colorlinks,linkcolor=blue,citecolor=blue,urlcolor=blue,hyperindex,driverfallback=dvipdfm]{hyperref}  \usepackage[T1]{fontenc} 

\def\ii{{\rm i}}  \def\ee{{\rm e}}
\def\me{m_{\rm e}}  
  
                                      \def\pb{{\bf p}}      \def\Rb{{\bf R}}  \def\rb{{\bf r}}      \def\vb{{\bf v}} 
    \def\zz{\hat{\bf z}}    \def\rr{\hat{\bf r}}        
  \def\red{\color{red}}    

\begin{document} 

\title{Complete Excitation of Discrete Quantum Systems by Single Free Electrons}


\author{F.~Javier~Garc\'{\i}a~de~Abajo}
\email{javier.garciadeabajo@nanophotonics.es} 
\affiliation{ICFO-Institut de Ciencies Fotoniques, The Barcelona Institute of Science and Technology, 08860 Castelldefels (Barcelona), Spain} 
\affiliation{ICREA-Instituci\'o Catalana de Recerca i Estudis Avan\c{c}ats, Passeig Llu\'{\i}s Companys 23, 08010 Barcelona, Spain}
\author{Eduardo~J.~C.~Dias}
\affiliation{ICFO-Institut de Ciencies Fotoniques, The Barcelona Institute of Science and Technology, 08860 Castelldefels (Barcelona), Spain}
\author{Valerio~Di~Giulio}
\affiliation{ICFO-Institut de Ciencies Fotoniques, The Barcelona Institute of Science and Technology, 08860 Castelldefels (Barcelona), Spain}


\begin{abstract}
We reveal a wealth of nonlinear and recoil effects in the interaction between individual low-energy electrons and samples comprising a discrete number of states. Adopting a quantum theoretical description of combined free-electron and two-level systems, we find a maximum achievable excitation probability of 100\%, which requires specific conditions relating to the coupling strength and the transition symmetry, as we illustrate through calculations for dipolar and quadrupolar modes. Strong recoil effects are observed when the kinetic energy of the probe lies close to the transition threshold, although the associated probability remains independent of the electron wave function even when fully accounting for nonlinear interactions with arbitrarily complex multilevel samples. Our work reveals the potential of free electrons to control localized excitations and delineates the boundaries of such control.
\end{abstract}

\maketitle 
\date{\today} 

\section{Introduction}

Free electron beams (e-beams) allow us to image material nanostructures and their excitations with an unsurpassed combination of space-energy resolution in the sub{\aa}ngstrom-meV domain thanks to a sustained series of advances in electron microscope instrumentation over the last decades \cite{NP98,BDK02,KLD14,LTH17,KDH19,HRK20,HHP19,paper369,YLG21}. In particular, electron energy-loss spectroscopy (EELS) is widely used to identify localized excitations and map their spatial distributions with atomic precision \cite{E96,E03,EB05,B06,paper149,KLD14,HNY18,HRK20,YLG21,paper371}, as exemplified by recent studies of photon confinement in optical cavities \cite{KLS20,WDS20,paper383}, atomic vibrations in thin layers \cite{HNY18,HRK20,YLG21} and molecules \cite{RAM16,HC18,HHP19}, and collective excitations such as phonon polaritons \cite{KLD14,LTH17,paper361,paper380} and plasmons \cite{BKW07,paper085,RB13,FWY14,paper369}.

At e-beam energies $>30\,$keV, typically employed in transmission electron microscopes to perform EELS analyses, the per-electron excitation probability of each individual mode in the specimen lies several orders of magnitude below unity. While such weak interaction is beneficial to grant us clean access into the nanoscale optical response over a wide spectral range ($10^{-3}-10^3\,$eV), a low excitation probability also implies that we operate in the linear regime, which is useless to track the ultrafast dynamics associated with a nonlinear behavior. This situation can be mitigated by resorting to less energetic probes like those available in low-energy electron microscopes \cite{R95,T19}. Indeed, individual $\lesssim100\,$eV electrons are predicted to generate multiple excitations of a single optical mode by appropriately adjusting the beam energy \cite{paper228}, while the onset of anharmonic response in this regime is expected to produce mode saturation and spectral shifts \cite{paper350}. In a different approach, femtosecond resolution is achieved in ultrafast electron microscopy by synchronizing laser and electron pulses in their arrival at the sampled structure \cite{BFZ09,paper151,FES15,PLQ15,KLS20,WDS20,HRF22}, a method that potentially enables the determination of nonlinear response functions with nanoscale resolution \cite{paper347}.

Many of the aforementioned studies focus on bosonic excitations (e.g., phonons \cite{KLD14,LTH17,paper361,paper380} and plasmons \cite{BKW07,paper085,RB13,FWY14,paper369}), which exhibit the characteristic linear response of harmonic oscillators, unless strong external fields are introduced to drive them beyond the parabolic potential region. In the opposite extreme, two- and few-level (fermionic) systems display a paradigmatic nonlinear behavior, whereby a given excitation can block subsequent ones. As an example, the discreteness of energy levels in nanographenes permeates their optical response and enables nonlinear interactions at the single-free-electron level \cite{paper350}. Nevertheless, fermionic excitations in systems such as atoms, molecules, and defect states in solids possess a weak transition strength that is essentially limited by the $f$-sum rule \cite{N1997} and, therefore, demands the use of low-energy electrons to yield measurable inelastic scattering signals.

Nonlinear effects open fundamental questions, such as whether an individual electron can produce a given excitation with 100\% probability, as well as the role of the electron wave function in determining that probability. In addition, we expect qualitatively different behavior between excitations of bosonic and fermionic character in the nonlinear regime. Because the probe energies required to reach a sizeable interaction strength are likely comparable to the transition energies, recoil effects are also anticipated to play an important role. These are relevant problems of the yet poorly explored terrain of nonlinear and recoil phenomena taking place during the interaction of free electrons with localized excitations.

In this Letter, we show that a free electron can excite a two-level system with 100\% probability, provided the transition symmetry and interaction strength meet specific conditions. Based on a quantum description of free electrons and localized excitations that rigorously incorporates nonlinear and recoil effects, we show that the excitation probability is independent of the electron wave function profile. Our calculations for bosonic and fermionic systems also demonstrate that recoil effects are irrelevant unless the electron energy is only a few times larger than the transition energy. Besides their fundamental interest, our results suggest a way to control localized excitations by means of free electrons, while establishing universal rules for the maximum achievable probability depending on the symmetry of the excitation and the electron-sample coupling strength.

\begin{figure}
\centering{\includegraphics[width=0.40\textwidth]{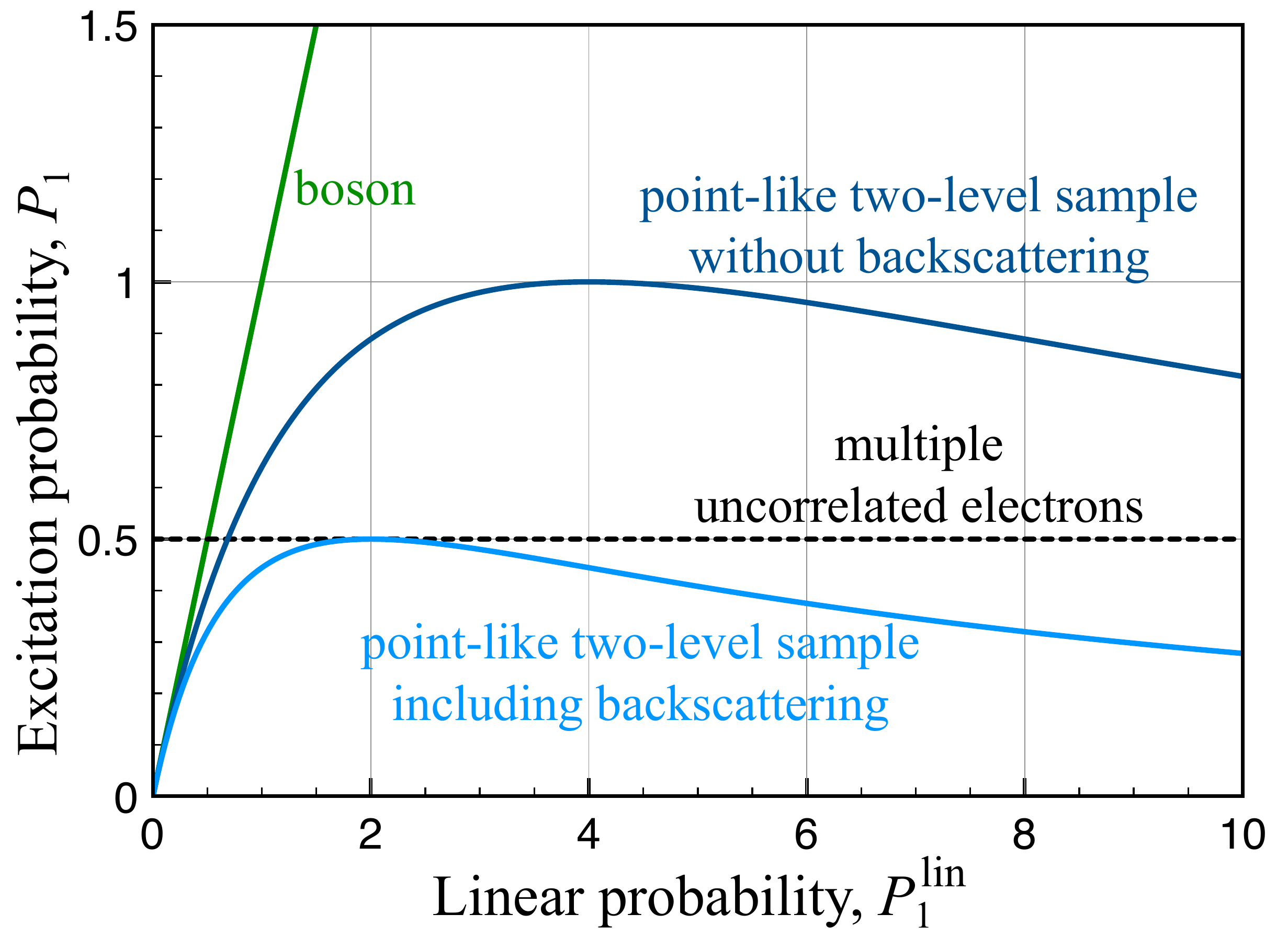}}
\caption{Nonlinear effects in the excitation of two-level systems by a single free electron. We represent the probability $P_1$ as a function of the first-order (linear) probability $P_1^{\rm lin}$ for a lossless point-like two-level system with and without inclusion of backscattering. The $P_1=P_1^{\rm lin}$ probability for a bosonic excitation is also shown for comparison. Many multiple uncorrelated electrons produce a probability of 1/2.}
\label{Fig1}
\end{figure}


\section{Nonperturbative excitation probability including recoil}

We consider a collimated e-beam focused down to a small lateral size at the region of interaction with the sampled structure, such that we can ignore its dynamics in a plane perpendicular to the beam direction $z$. We further assume a nonlossy specimen comprising a discrete set of states $|j\rangle$ of energies $\hbar\omega_j$ and initially placed in the the ground state $|0\rangle$. The Hamiltonian of the combined electron-sample system can be written
\begin{align}
\hat{\mathcal{H}}=&\hbar\int\!dq\;\varepsilon_q\,|q\rangle\langle q|+\hbar\sum_j\omega_j|j\rangle\langle j|
\nonumber\\
&+\hbar\int\!dq\int\!dq'\sum_{jj'}G_{qj,q'j'}|qj\rangle\langle q'j'|,
\nonumber
\end{align}
where the electron is represented by orthonormal momentum states $|q\rangle$ of energies $\hbar\varepsilon_q$, whereas $G_{qj,q'j'}$ are electron-sample coupling coefficients. Expanding the wave function of the combined system as
\begin{align}
|\Psi(t)\rangle=\int\!dq\sum_j\ee^{-\ii(\varepsilon_q+\omega_j)t}\alpha_{qj}(t)\,|qj\rangle,
\nonumber
\end{align}
inserting it into the Schr\"odinder equation $\hat{\mathcal{H}}|\Psi(t)\rangle=\ii\hbar|\dot{\Psi}(t)\rangle$, and adopting the initial conditions $\alpha_{qj}(-\infty)=\alpha_q^0\,\delta_{j0}$ [i.e., with the specimen in the ground state $j=0$ and an incident electron wave function $\psi^0(z,t)\propto\int\!dq\,\alpha_q^0\,\ee^{\ii(qz-\varepsilon_qt)}$], we find the post-interaction solution (see Appendix)
\begin{align}
\alpha_{qj}(\infty)&=\alpha_q^0\,\delta_{j0}-2\pi\ii\frac{M_{q\tilde{q}_j,j}}{v_{\tilde{q}_j}}\;\alpha_{\tilde{q}_j}^0, \label{alphainfty}
\end{align}
where the coefficients $M_{qq',j}$ are independent of the incident electron state and satisfy to the self-consistent Lippmann–Schwinger \cite{M1966} relation
\begin{align}
M_{qq',j}=G_{qj,q'0}-\!\int\!dq''\sum_{j'}\frac{G_{qj,q''j'}M_{q''q',j'}}{\varepsilon_{q''q'}+\omega_{j'0}-\ii0^+}
\label{Mqqj}
\end{align}
with $\varepsilon_{qq'}=\varepsilon_{q}-\varepsilon_{q'}$ and $\omega_{jj'}=\omega_j-\omega_{j'}$. Here, $v_q=d\varepsilon_q/dq$ is the group velocity of the $q$ electron component, while $\tilde{q}_j$ is implicitly defined by $\varepsilon_{\tilde{q}_j}=\varepsilon_q+\omega_{j0}$ with $\tilde{q}_j>0$ (i.e., $\alpha_{\tilde{q}_j}^0$ only contains forward propagating components).

We are interested in the probability $P_j$ for a sampled system initially prepared in its ground state $|0\rangle$ to be left in state $|j\rangle$ after the interaction has taken place. We thus write $P_j=\int\!dq\;\big|\alpha_{qj}(\infty)\big|^2$, which upon insertion of Eq.\ (\ref{alphainfty}), leads to a decomposition of the probability in incident-momentum components according to (see Appendix)
\begin{align}
P_j=\int_{q_{\rm min}^j}^\infty\!dq\,\big|\alpha_q^0\big|^2\,P_{q,j},
\label{Pj}
\end{align}
where
\begin{align}
&P_{q,j}=\frac{4\pi^2}{v_{q_j}v_q}\left(\left|M_{q_j,q,j}\right|^2+\big|M_{-q_j,q,j}\big|^2\right)
\label{Pqj}
\end{align}
for excited states $j\neq0$. Here, the final electron wave vector $q_j>0$ is defined through $\varepsilon_{q_j}=\varepsilon_q-\omega_{j0}$, and a minimun incident wave vector $q_{\rm min}^j$ is imposed by the threshold excitation condition $\varepsilon_{q_{\rm min}^j}=\omega_{j0}$. The first and second terms in Eq.\ (\ref{Pqj}) correspond to the contributions of forward and backward electron scattering (i.e., final wave vectors $q_j$ and $-q_j$, respectively). This result reveals a trivial role of the incident electron wave function: each initial wave vector component contributes to the excitation probability in proportion to $\big|\alpha_q^0\big|^2$ [see Eq.\ (\ref{Pj})], with no dependence on the phase of $\alpha_q^0$ [i.e., on the profile of the incident wave function $\psi^0(z,t)$]. We remark that this conclusion is derived from a nonperturbative formalism that rigorously accounts for nonlinear and recoil effects.

Because the energy spread of the incident beam plays a trivial role, we limit our discussion to monochromatic electrons of energy $\hbar\varepsilon_{q_0}$ with $\big|\alpha_q^0\big|^2=\delta(q-q_0)$, so that the excitation probability reduces to $P_j=P_{q_0,j}$, subject to the condition $q_0>q_{\rm min}^j$. In addition, we focus on two-level systems, although the present formalism can be readily applied to multilevel configurations. We thus concentrate on the excitation probability $P_1$ and also consider the linear probability $P_1^{\rm lin}$ for reference, obtained from Eqs.\ (\ref{Mqqj}) and (\ref{Pqj}) by neglecting the integral term in the former (see Appendix).

\section{Point-like interaction limit}

As a first tutorial step, we obtain a closed-form solution when the interaction is localized to just one point, so that the coupling coefficients $G_{qj,q'j'}$ are independent of $q$ and $q'$. Then, the excitation probability reduces to (see Appendix)
\begin{align}
P_1=\frac{P_1^{\rm lin}}{\left(1+P_1^{\rm lin}/2\right)^2},
\nonumber
\end{align}
which presents a single maximum $P_1=1/2$ as a function of the linear probability at $P_1^{\rm lin}=2$, as shown in Fig.\ \ref{Fig1}. Only for this case, we include backscattering in the linear probability. This result already reveals that maximum excitation is only achieved for a very specific value of the coupling coefficient or, alternatively, $P_1^{\rm lin}$. Interestingly, the presence of two inelastic channels (forward and backward scattering) limits the maximum probability to 50\%. Indeed, if we disregard backscattering, which should be reasonable for energetic electrons, a similar analysis leads to (see Appendix)
\begin{align}
P_1=\frac{P_1^{\rm lin}}{\left(1+P_1^{\rm lin}/4\right)^2},
\nonumber
\end{align}
whose maximum value is now $P_1=1$, obtained at $P_1^{\rm lin}=4$. For comparison, we show the $P_1=P_1^{\rm lin}$  line corresponding to a bosonic mode, and obviously, all of these results are in mutual agreement in the $P_1^{\rm lin}\ll1$ limit. Incidentally, the average population of the excited state in a two-level system interacting with many multiple uncorrelated electrons is $1/2$ \cite{paper371}.

\begin{figure*}
\centering{\includegraphics[width=0.9\textwidth]{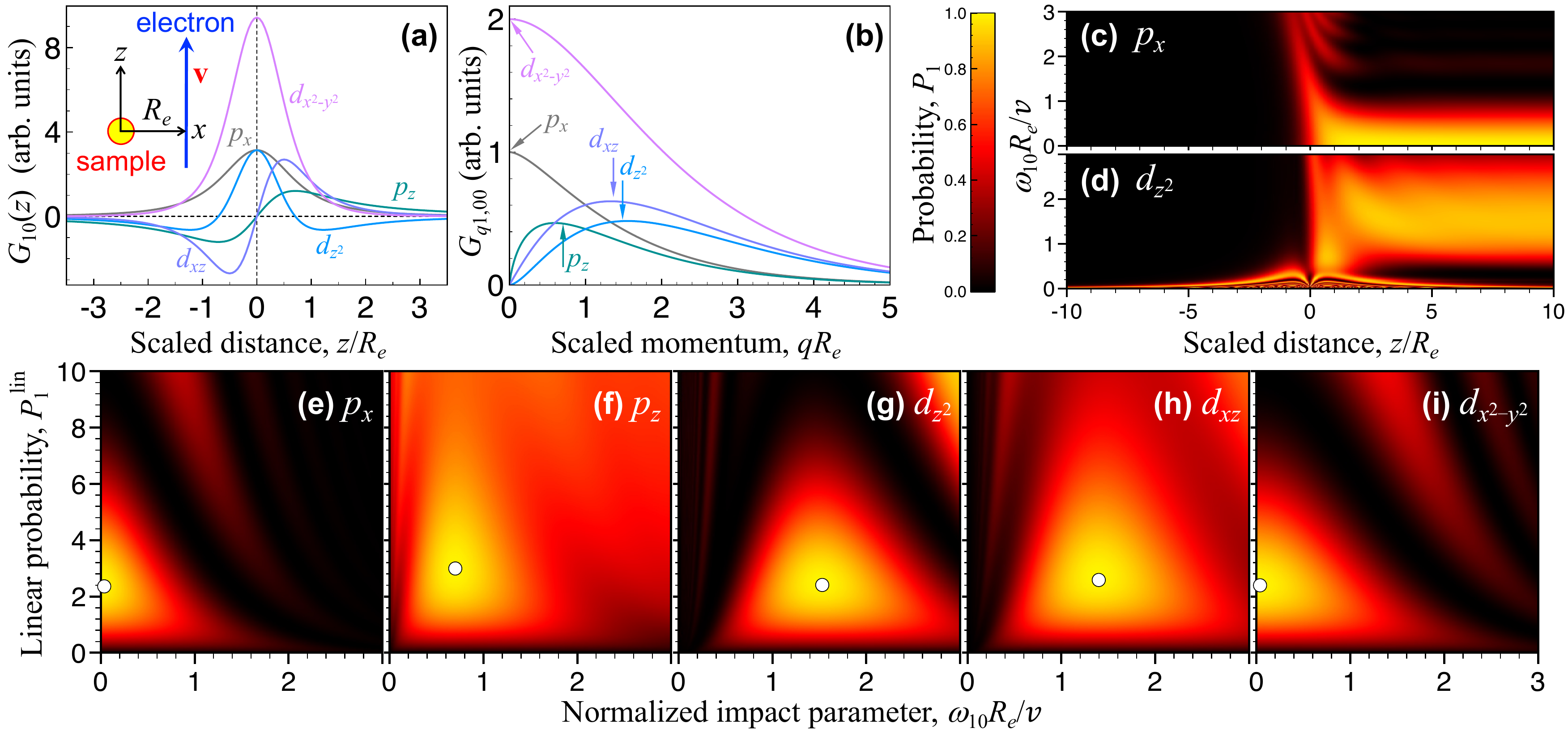}}
\caption{Excitation of two-level systems by a single free electron. We consider the configuration illustrated in the inset of panel (a) and study the excitation probability for different transition symmetries. (a),(b) Electron-sample interaction coefficients $G_{10}(z)$ and $G_{q1,00}$ in the real-space (a) and momentum-space (b) representations, respectively, for dipolar ($p_x$ and $p_z$) and quadrupolar ($d_{z^2}$, $d_{xz}$, and $d_{x^2-y^2}$) excitations with different angular symmetries. (c),(d) Evolution of the excited state occupation as a function of position along the electron trajectory $z$ and impact parameter $R_e$ for two selected dipolar and quadrupolar excitation symmetries (see labels). (e)-(i) Dependence of the post-interaction excitation probability (at $z\rightarrow\infty$) as a function of impact parameter $R_e$ and linear probability $P_1^{\rm lin}$ for all nonvanishing dipolar and quadrupolar excitation symmetries under the investigated beam-sample configuration. The probability reaches 100\% at the positions indicated by the white dots in (e)-(i). We take fixed values of $P_1^{\rm lin}=2.5$ and 1.5 in (c) and (d), respectively.}
\label{Fig2}
\end{figure*}

\section{Nonlinear e-beam excitation without recoil}

As we show below, recoil effects can be neglected if the electron energy exceeds several times the transition energy. We can then linearize the electron energy difference as $\varepsilon_{qq'}\approx(q-q')v$, where $v$ is the electron velocity. Considering a small sampled system, whose interaction with low-energy electrons can be described through the Coulomb potential, we find the associated coupling coefficients to only depend on the wave vector difference $q-q'$ and take the form $G_{qj,0j'}\propto\big({\rm sign}\{q\}\big)^\sigma |q|^lK_m(|q|R_e)$, where $R_e$ is the beam-sample distance, $(l,m)$ are the angular momentum numbers associated with the excitation symmetry, $\sigma$ takes values of 0 or 1, and a constant of proportionality depending on the details of the system is taken to be absorbed in $P_1^{\rm lin}$. In particular, we consider excitations of dipolar [$p_x$ and $p_z$, corresponding to $(l,m,\sigma)=(1,1,0)$ and $(1,0,1)$, respectively] and quadrupolar [$d_{z_2}$, $d_{xz}$, and $d_{x^2-y^2}$, corresponding to $(2,0,0)$, $(2,1,1)$, and $(2,2,0)$] character, with a geometrical configuration as shown in the inset of Fig.\ \ref{Fig2}(a) (see Appendix for details, and Fig.\ \ref{Fig2}(b) for the associated momentum-space coupling coefficients).

Under these conditions, the wave function of the system admits the form (see Appendix)
\begin{align}
\langle z|\Psi(t)\rangle=\psi^0(z,t) \sum_j f_j(z)\,\ee^{-\ii\omega_{j0}(z/v-t)}\,\ee^{-\ii\omega_jt}\,|j\rangle, \nonumber
\end{align}
where the space-dependent functions $f_j(z)$ evolve as
\begin{align}
\frac{d\,f_j(z)}{dz}&=-\frac{\ii}{v}\sum_{j'} G_{jj'}(z)\;\ee^{\ii\omega_{jj'}z/v}\,f_{j'}(z), \label{df}
\end{align}
and we introduce real-space coupling coefficients $G_{jj'}(z)=\int\!dq\;G_{qj,0j'}\,\ee^{\ii qz}$ [see Fig.\ \ref{Fig2}(a)]. Finally, the excitation probability is simply given by $P_j=\big|f_j(\infty)\big|^2$, while the linear limit reduces to $P_j^{\rm lin}=(4\pi^2/v^2)\big|G_{0j,\omega_{j0}/v,0}\big|^2$ for $j\neq0$ (see Appendix).

We numerically integrate Eq.\ (\ref{df}) for two-level systems with the excitation symmetries noted above to obtain the universal plots of $P_1$ presented in Figs.\ \ref{Fig2}(e)-\ref{Fig2}(i) as a function of the dimensionless parameters $\omega_{10}R_e/v$ and $P_1^{\rm lin}$. Remarkably, $P_1$ reaches a single maximum of 100\% at a specific $(R_e,P_1^{\rm lin})$ point [white dots in Figs.\ \ref{Fig2}(e)-\ref{Fig2}(i)]. The position of this maximum occurs at values of $P_1^{\rm lin}$ that are in the range of those obtained in the point-interaction limit (Fig.\ \ref{Fig1}), while the impact parameter $R_e$ lies close to the stationary points of $G_{q1,00}$ as a function of $qR_e$ for a wave vector transfer $q=\omega_{10}/v$ determined by the nonrecoil approximation [cf. the maxima of the curves in Fig.\ \ref{Fig2}(b) and the abscissas of the white dots in Figs.\ \ref{Fig2}(e)-\ref{Fig2}(i)]. Two of the studied symmetries have this maximum at $R_e=0$, accompanied by a lack of any zeros in the real-space profile of the corresponding coupling coefficients [Fig.\ \ref{Fig2}(a)], in contrast to the other excitations under consideration. Incidentally, $P_1$ presents multiple maxima as we move along $R_e$ for fixed $P_1^{\rm lin}$, the magnitudes of which decrease with increasing impact parameter. This is the result of a complex evolution of the position-dependent probability $|f_1(z)|^2$ along the electron path, which exhibits oscillations before reaching an asymptotic value of $P_1$ at large $z$ [see examples of this dynamics in Figs.\ \ref{Fig2}(c),\ref{Fig2}(d)].

\begin{figure}
\centering{\includegraphics[width=0.45\textwidth]{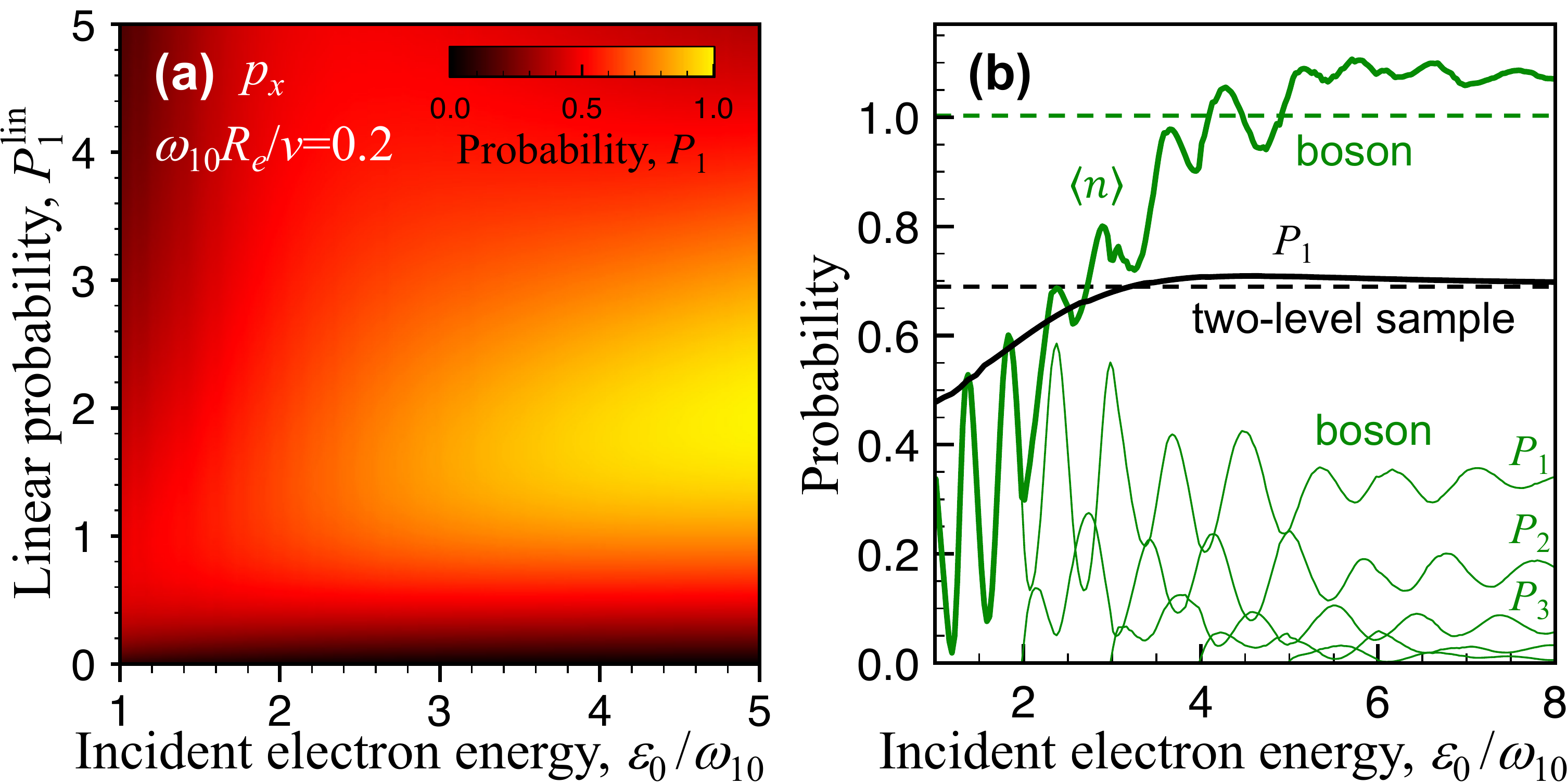}}
\caption{Recoil effects in near-edge excitation. (a) Excitation probability $P_1$ for a two-level sample as a function of incident electron energy (horizontal axis, normalized to the excitation energy $\hbar\omega_{10}$) and linear excitation probability $P_1^{\rm lin}$. (b) Probability extracted from panel (a) for fixed $P_1^{\rm lin}=1$ (black curve) compared with the excitation probability for a bosonic mode with the same $0\rightarrow1$ matrix element (green curves, comprising a decomposition in the contributions $P_n$ of different Fock states $|n\rangle$, as well as the final average population $\langle n\rangle=\sum_{n=1}^\infty n\,P_n$). Dashed curves inticate the $\varepsilon_0\gg\omega_{10}$ nonrecoil limit. We consider dipolar excitations of $p_x$ symmetry and a normalized impact parameter $\omega_{10}R_e/v=0.2$ in all cases.}
\label{Fig3}
\end{figure}

\section{Effect of electron recoil}

The solution of Eqs.\ (\ref{Mqqj}) and (\ref{Pqj}) for a two-level system produces an excitation probability that is substantially reduced with respect to the nonrecoil limit when the incident electron energy $\hbar\varepsilon_0$ approaches the excitation threshold $\hbar\omega_{10}$. We illustrate this effect in Fig.\ \ref{Fig3}(a) for $\omega_{10}R_e/v=0.2$ and $p_x$ transition symmetry over a wide range of coupling strengths (vertical axis), but we find that this conclusion is general upon extensive numerical inspection of different $R_e$ values. The nonrecoil result is however recovered when $\varepsilon_0$ is just a few times $\omega_{10}$. A similar effect of recoil is observed in the excitation of a bosonic mode [Fig.\ \ref{Fig3}(b)], although the interplay between different Fock states $|n\rangle$ leads to a more complex evolution characterized by sharp oscillations in both the total excitation probability and the partial contribution coming from each $|n\rangle$ state. These oscillations are attenuated as $\varepsilon_0$ increases, leading to a Poissonian distribution \cite{paper228,paper339} (see also Appendix).

\section{CONCLUSIONS}

In conclusion, a wealth of phenomena unfolds from the interaction between free electrons and few-level systems. In particular, we have shown that achieving complete excitation of a single transition in a specimen by an individual free electron is not simply a matter of increasing the interaction strength, but it also requires a specific balance that depends on the symmetry of the excited mode. In addition, the excitation probability is independent of the electron wave function profile even when fully accounting for nonlinear and recoil effects. Low-energy electrons in the $<100\,$eV range are promising to explore these effects, as they can generate multiple excitations of a single plasmon mode in atomically thin nanostructures \cite{paper228}. Excitons in two-dimensional materials \cite{TLM15} offer a potentially practical candidate to study the iteration of free electrons with few-level systems, while defect states in those materials, already explored with tunneling microscopes \cite{paper354}, are robust two-level systems that could be investigated with low-energy electrons in a reflection configuration. Free-electron interaction with diluted atomic or molecular gases could also serve as a platform to study the coupling strength, while incipient electron microscopy studies on optical atomic lattices and condensates \cite{GWR08} could be extended to measure inelastic scattering and explore the physics portrayed in the present work.

\section*{ACKNOWLEDGMENTS}
This work has been supported in part by the European Research Council (Advanced Grant 789104-eNANO), the European Commission (Horizon 2020 Grants 101017720 FET-Proactive EBEAM and 964591-SMART-electron), the Spanish MICINN (PID2020-112625GB-I00 and Severo Ochoa CEX2019-000910-S), the Catalan CERCA Program, and Fundaci\'{o}s Cellex and Mir-Puig.

\section*{APPENDIX}

\appendix 



We provide a self-contained derivation of the theory used in the main text to describe the interaction between a single collimated free electron and a system comprising a discrete set of quantum states. The excitation probabilities are shown to only depend on the spectral distribution of the incident electron, but not on the phase and shape of its wave function, even when recoil, nonlinear, and relativistic effects are accounted for in a rigorous manner. We also present explicit expressions for the coupling coefficients between the electron and selected excitations with well-defined multipolar symmetry in the nonrelativistic limit. A simpler solution to the interaction problem is further elaborated within the nonrecoil approximation. Finally, we offer details of a numerical implementation using nonrelativistic kinematics, but fully incorporating recoil and nonlinear effects.

\section{Free-electron interaction with a discrete-level system beyond the linear nonrecoil regime}
\label{sec1}

We study the interaction between a collimated free electron and a sample comprising a discrete set of states $|j\rangle$ of energies $\hbar\omega_j$ by adopting the following assumptions:
\begin{enumerate}[\it\red {\rm(}i\rm{)} ---]
\item {\it Longitudinal motion.---}The electron is tightly focused in the transverse plane (perpendicular to the propagation direction $z$) down to a small region in which the interaction with the sample is approximately homogeneous [i.e., independent of the transverse coordinates $\Rb=(x,y)$ across the electron beam (e-beam)]. In addition, the transverse electron wave function remains nearly unchanged during the interaction time, so that we can dismiss it as well as any change in the transverse electron energy.
\item {\it Nonlossy specimen.---}Inelastic decay of the excited sample states plays a negligible role during the interaction time.
\end{enumerate}
Item {\it \red {\rm(}i\rm{)}} allows us to describe the electron in terms of a basis set of momentum states $|q\rangle$ of energies $\hbar\varepsilon_q$ (with $\varepsilon_q=c\sqrt{(\me c/\hbar)^2+q^2}$ or $\varepsilon_q=\hbar q^2/2\me$ within relativistic or nonrelativistic kinematics, respectively) and wave functions $\langle z|q\rangle=\ee^{\ii qz}/\sqrt{2\pi}$ satisfying the orthonormalization relation $\langle q|q'\rangle=\delta(q-q')$. In addition, point {\it \red {\rm(}ii\rm{)}} permits describing the evolution of the system by solving the Schr\"odinger equation with the total Hamiltonian
\begin{align}
&\hat{\mathcal{H}}=\hbar\int\!dq\;\varepsilon_q\,|q\rangle\langle q|
+\hbar\sum_j\omega_j|j\rangle\langle j| \nonumber\\
&+\hbar\int\!dq\int\!dq'\sum_{jj'}G_{qj,q'j'}|qj\rangle\langle q'j'|,
\label{H}
\end{align}
where $G_{qj,q'j'}=\hbar^{-1}\langle qj|\hat{\mathcal{H}}|q'j'\rangle$ are electron-sample coupling coefficients.

We proceed by writing the wave function of the combined electron-sample system as
\begin{align}
|\Psi(t)\rangle=\int\!dq\sum_j\ee^{-\ii(\varepsilon_q+\omega_j)t}\alpha_{qj}(t)\,|qj\rangle,
\label{psi}
\end{align}
which, upon insertion into $\hat{\mathcal{H}}|\Psi(t)\rangle=\ii\hbar|\dot{\Psi}(t)\rangle$, leads to the equation of motion
\begin{align}
\dot{\alpha}_{qj}(t)&=-\ii\int\!dq'\sum_{j'} G_{qj,q'j'} \,\alpha_{q'j'}(t)\, \ee^{\ii(\varepsilon_{qq'}+\omega_{jj'})t} \label{dadt}
\end{align}
for the expansion coefficients in Eq.\ (\ref{psi}), where we use the compact notation $\varepsilon_{qq'}=\varepsilon_q-\varepsilon_{q'}$ and $\omega_{jj'}=\omega_j-\omega_{j'}$. Obviously, the Hamiltonian in Eq.\ (\ref{H}) is Hermitian, so the normalization condition $\int\!dq\sum_j|\alpha_{qj}|^2=1$ is maintained during time propagation.

We are interested in studying the probability
\begin{align}
P_j=\int\!dq\;\big|\alpha_{qj}(\infty)\big|^2
\label{Pj}
\end{align}
that a sample initially prepared in its ground state $|0\rangle$ is left in a state $|j\rangle$ after the interaction has taken place. Consequently, we set $\alpha_{qj}(-\infty)=\alpha_q^0\,\delta_{j0}$, where the coefficients $\alpha_q^0$ define the incident electron wave function
\begin{align}
\psi^0(z,t)=\int\!dq\; \alpha_q^0\; \frac{\ee^{\ii(qz-\varepsilon_qt)}}{\sqrt{2\pi}},
\label{psi0}
\end{align}
whose normalization ($\int dz\; \big|\psi^0(z,t)\big|^2=1$) imposes the condition $\int\!dq\;\big|\alpha_q^0\big|^2=1$.

At this point, we anticipate a solution of the form
\begin{align}
\alpha_{qj}(t)=\alpha_q^0\,\delta_{j0}-\int\!dq'\; M_{qq',j}\;\alpha_{q'}^0\;\frac{\ee^{\ii(\varepsilon_{qq'}+\omega_{j0})t}}{\varepsilon_{qq'}+\omega_{j0}-\ii0^+},
\label{aM}
\end{align}
which is suggested by iteratively integrating Eq.\ (\ref{dadt}) in a perturbation-theory approach. Indeed, inserting Eq.\ (\ref{aM}) into Eq.\ (\ref{dadt}) and integrating over time, we find the self-consistent Lippmann–Schwinger equation \cite{M1966}
\begin{align}
M_{qq',j}=G_{qj,q'0}-\int\!dq''\sum_{j'}\frac{G_{qj,q''j'}M_{q''q',j'}}{\varepsilon_{q''q'}+\omega_{j'0}-\ii0^+}
\label{NGGN}
\end{align}
for the coefficients in Eq.\ (\ref{aM}), which are independent of the incident electron state. Now, applying the identity $\exp(\ii\theta t)/(\theta-\ii0^+)\xrightarrow[t\to\infty]{}2\pi\ii\delta(\theta)$ with $\theta=\varepsilon_{qq'}+\omega_{j0}$ to Eq.\ (\ref{aM}), we find the post-interaction solution
\begin{align}
\alpha_{qj}(\infty)&=\alpha_q^0\,\delta_{j0}-2\pi\ii\int\!dq'\; M_{qq',j}\;\alpha_{q'}^0\;\delta(\varepsilon_{qq'}+\omega_{j0}) \nonumber\\
&=\alpha_q^0\,\delta_{j0}-2\pi\ii\frac{M_{q\tilde{q}_j,j}}{v_{\tilde{q}_j}}\;\alpha_{\tilde{q}_j}^0,
\nonumber
\end{align}
where $v_q=d\varepsilon_q/dq$ is the group velocity of the $q$ electron component, and the rightmost expression is obtained by manipulating the $\delta$-function as $\delta(\varepsilon_{qq'}+\omega_{j0})=\big[\delta(q'-\tilde{q}_j)+\delta(q'+\tilde{q}_j)\big]/v_{\tilde{q}_j}$ with $\tilde{q}_j$ implicitly defined by $\varepsilon_{\tilde{q}_j}=\varepsilon_q+\omega_{j0}$. Here, we consider an incident electron that only contains $\tilde{q}_j>0$ components in $\alpha_{\tilde{q}_j}^0$ (i.e., it moves toward increasing $z$), but the final state can receive both $q>0$ (forward scattering) and $q<0$ (backscattering) contributions. Finally, inserting this result into Eq.\ (\ref{Pj}), using the identity $v_{\tilde{q}_j}d\tilde{q}_j=v_qdq$, and changing the variable of integration as $\tilde{q}_j\rightarrow q$, the post-interaction occupation probability of level $j$ reduces to
\begin{align}
P_j&=\int_{q_{\rm min}^j}^\infty\!\frac{dq}{v_{q_j}v_q}\label{Pjfinal}\\
&\times\left[\left|\left(\delta_{j0}\,v_q-2\pi\ii\,M_{q_j,q,j}\right)\right|^2+4\pi^2\,\big|M_{-q_j,q,j}\big|^2\right]\;\big|\alpha_q^0\big|^2,
\nonumber
\end{align}
where $q_j$ depends on $q$ and is now defined by $\varepsilon_{q_j}=\varepsilon_q-\omega_{j0}$ (i.e., $q_j=(\me c/\hbar)\big[\big(\sqrt{1+(\hbar q/\me c)^2}-\hbar\omega_{j0}/\me c^2\big)^2-1\big]^{1/2}$ or $q_j=\sqrt{q^2-2\me\omega_{j0}/\hbar}$ within relativisitic or nonrelativistic kinematics, respectively); a wave vector threshold $q_{\rm min}^j$ is imposed by the minimum electron energy capable of exciting the $j$ level in the sample (i.e., $\varepsilon_{q_{\rm min}^j}=\omega_{j0}$); and the first and second terms inside the square brackets account for forward and backward electron scattering contributions with final electron wave vectors $q_j$ and $-q_j$, respectively. The result embodied in Eq.\ (\ref{Pjfinal}) is general within the approximations in points {\it \red {\rm(}i\rm{)}} and {\it \red {\rm(}ii\rm{)}} above, and it shows a trivial dependence on the incident electron wave function: the contribution of each incident wave vector component $q$ to the excitation probability is weighted by its strength $\big|\alpha_q^0\big|^2$, so the phase in $\alpha_q^0$ (i.e., the wave function profile) does not play any role at all. {\it We stress that the present derivation demonstrates that this conclusion is maintained even when rigorously accounting for nonlinear electron-sample interactions and recoil effects.}

\subsection{Solution for monochromatic electrons}

Because the energy spread of the incident electron plays a trivial role, we can limit our discussion to monochromatic electrons of incident energy $\hbar\varepsilon_{q_0}$, characterized by $\big|\alpha_q^0\big|^2=\delta(q-q_0)$. Then, the excitation probability reduces to
\begin{align}
P_j=&\frac{\Theta(\epsilon_{q_0}-\omega_{j0})}{v_{q_j}v_{q_0}} \label{Pjfinalfinal}\\
&\times\left[\left|\delta_{j0}\,v_{q_0}-2\pi\ii\,M_{q_j,j}\right|^2
+4\pi^2\,\big|M_{-q_j,j}\big|^2\right],
\nonumber
\end{align}
where the coefficients $M_{qj}\equiv M_{qq_0,j}$ satisfy the self-consistent equation
\begin{align}
M_{qj}=G_{qj,q_00}-\int\!dq'\sum_{j'}\frac{G_{qj,q'j'}M_{q'j'}}{\varepsilon_{q'q_0}+\omega_{j'0}-\ii0^+},
\label{Njfinal}
\end{align}
which we write directly from Eq.\ (\ref{NGGN}). We remark again that the first and second terms inside the square brackets of Eq.\ (\ref{Pjfinalfinal}) correspond to forward and backward electron scattering contributions, with associated final electron wave vectors determined as illustrated in the following sketch:

\begin{centering} \includegraphics[width=0.32\textwidth]{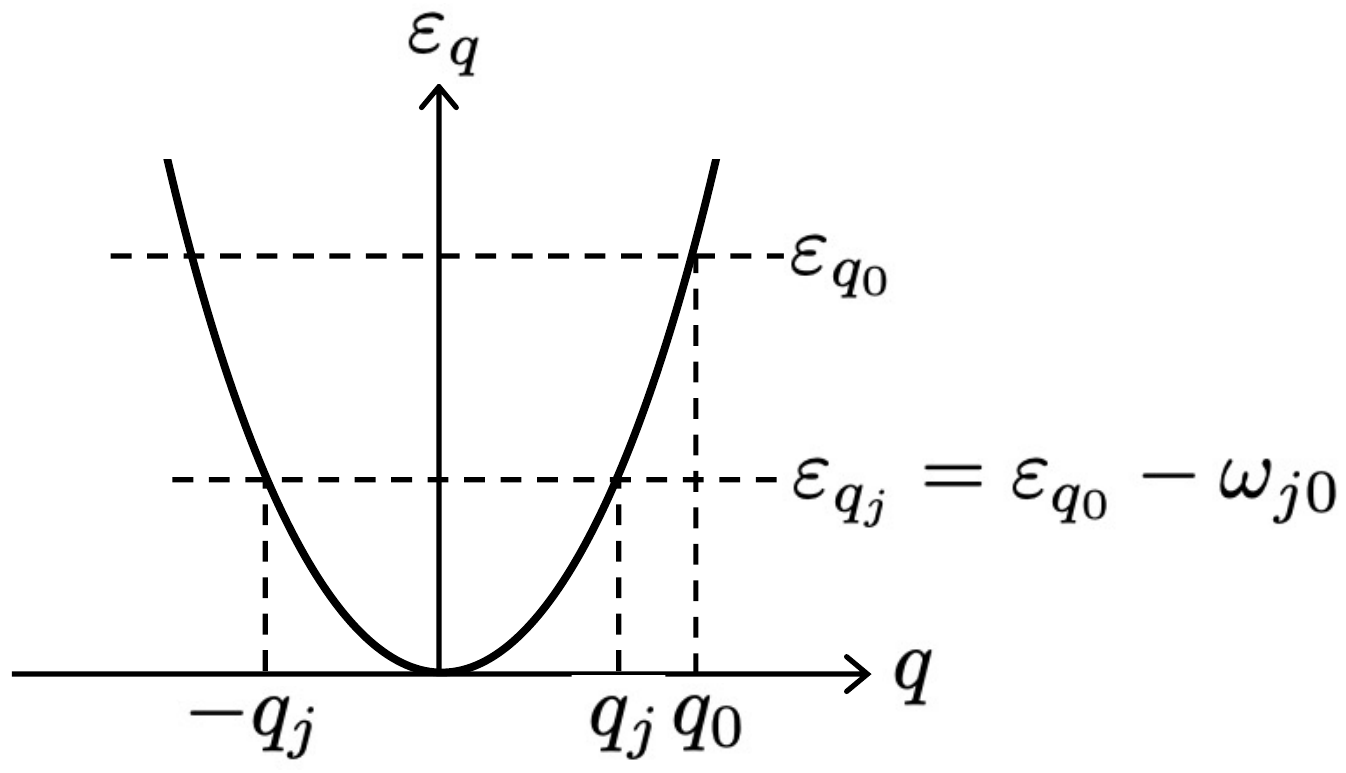} \par\end{centering}
\noindent Obviously, backscattering is generally negligible, unless the electron energy is close to the threshold $\hbar\omega_{j0}$. For reference, we also consider the first-order (linear) excitation probabilities
\begin{align}
P_j^{\rm lin}=\frac{4\pi^2}{v_{q_0}v_{q_j}}\left(\left|G_{q_jj,q_00}\right|^2+\left|G_{-q_jj,q_00}\right|^2\right),
\label{P1lin}
\end{align}
derived from Eq.\ (\ref{Pjfinalfinal}) for $j\neq0$ by neglecting the integral term in Eq.\ (\ref{Njfinal}) (i.e., setting $M_{qj}=G_{qj,q_00}$).

\subsection{Point-like excitation limit with backscattering}

Interestingly, exact solutions can be obtained when the interaction is localized to just a single point at $z=0$. Indeed, a point-like interaction is characterized by coupling coefficients $G_{qj,q'j'}\equiv G_{jj'}$ that are independent of $q$ and $q'$ [i.e., giving rise to a real-space interaction coefficients of the form $G_{jj}(z)\propto\delta(z)$ according to Eq.\ (\ref{Gjjz}) in Sec.\ \ref{exact} below]. In this scenario, further adopting nonrelativisitc kinematics, so that $\varepsilon_q$ is quadratic in $q$, we find that Eq.\ (\ref{Njfinal}) reduces to $M_j=G_{j0}-2\pi\ii\sum_{j'}G_{jj'}\,M_{j'}/v_{q_{j'}}$ \cite{noteonGqqbis}, whose solutions $M_j\equiv M_{qj}$ are also independent of $q$. In particular, for a two-level system comprising the states $j=0$ and 1, the only nonzero coupling elements are $G_{10}=G^*_{01}$, so combining the solution for $M_1$ with Eq.\ (\ref{Pjfinalfinal}), we find the excitation probability
\begin{align}
P_1=\frac{P_1^{\rm lin}}{\left(1+P_1^{\rm lin}/2\right)^2}.
\nonumber
\end{align}
We note that forward and backward scattering events contribute identically to the linear probability under the assumption of a $q$-independent coupling, which leads to $P_1^{\rm lin}=8\pi^2\big|G_{10}\big|^2/v_{q_0}v_{q_1}$ [see Eq.\ (\ref{P1lin})]. Then, a single absolute maximum $P_1=1/2$ is obtained for $P_1^{\rm lin}=2$. Reassuringly, this approximation maintains the overall probability $P_0+P_1=1$.

\subsection{Point-like excitation limit without backscattering}

When considering energetic electrons, a point-like scatterer is still described by $q$-independent coupling coefficients for relatively small changes in wave vector. However, backscattering (i.e., with $q<0$ and $q'>0$) involves large momentum transfers, for which $G_{qj,q'j'}$ becomes negligible at large incident electron energy. Then, we obtain $q$-independent coefficients $M_j\equiv M_{qj}=G_{j0}-\ii\pi\sum_{j'}G_{jj'}\,M_{j'}/v_{q_{j'}}$ for $q>0$ (notice the absence of a factor of 2 with respect to the result including backscattering), while $M_{qj}=0$ for $q<0$. Finally, we find the solution
\begin{align}
P_1=\frac{P_1^{\rm lin}}{\left(1+P_1^{\rm lin}/4\right)^2}
\nonumber
\end{align}
with $P_1^{\rm lin}=4\pi^2\big|G_{10}\big|^2/v_{q_0}v_{q_1}$ [i.e., also neglecting backscattering in Eq.\ (\ref{P1lin})], which now presents a single absolute maximum $P_1=1$ when $P_1^{\rm lin}=4$. In addition, the condition $P_0+P_1=1$ is also verified.

\section{Numerical implementation}
\label{exact}

The coupling coefficients $G_{qj,q'j'}$ are generally smooth functions of $q$ and $q'$, so we can solve Eq.\ (\ref{Njfinal}) by discretizing $q$ through a set of $2N+1$ points $p_l=q_0+lh$, each of them representing an interval $p_l-h/2<q<p_l+h/2$ labeled by $l=-N,\cdots,N$. Consecutive points are separated by a constant spacing $h=\Delta/(2N+1)$, and together they cover a finite range of size $\Delta$ centered at the incident electron wave vector $q_0$. We can then recast Eq.\ (\ref{Njfinal}) into the finite matrix equation
\begin{align}
M_j=g_j-\sum_{j'}S_{jj'}\cdot M_{j'},
\label{Mjblocks}
\end{align}
where $M_j$ and $g_j$ denote vectors of $2N+1$ components $M_{j,l}\equiv M_{q_lj}$ and $g_{j,l}\equiv G_{p_lj,q_00}$, respectively, whereas $S_{jj'}=G_{jj'}\cdot\Delta_{j'}$ are $(2N+1)\times(2N+1)$ square matrices defined in terms of the coupling matrices $G_{jj',ll'}\equiv G_{p_lj,p_{l'}j'}$ and the diagonal matrix
\begin{align}
\Delta_{j,ll'}=\delta_{ll'}\int_{p_l-h/2}^{p_l+h/2} \frac{dq}{\varepsilon_{qq_0}+\omega_{j0}-\ii0^+}.
\nonumber
\end{align}
In practice, a strong electron-sample interaction is likely involving relatively small electron velocities, and therefore, we can work in the nonrelativistic limit and write  $\varepsilon_q=\hbar q^2/2\me$ and $v_q=\hbar q/\me$. The matrix elements of $\Delta_j$ are then given by the closed-form expressions
\begin{widetext}
\begin{align}
\Delta_{j,ll'}=\delta_{ll'}\;\frac{\me}{\hbar|q_j|}\times
\left\{\begin{array}{ll} \log\left|\big[p_l^2-(|q_j|-h/2)^2\big]\big/\big[p_l^2-(|q_j|+h/2)^2\big]\right|+\ii\pi\,\theta_{jl}, &\quad\quad \varepsilon_{q_0}>\omega_{j0}, \\ & \\
2\left(\tan^{-1}\big[(p_l+h/2)/|q_j|\big]-\tan^{-1}\big[(p_l-h/2)/|q_j|\big]\right), &\quad\quad \varepsilon_{q_0}<\omega_{j0}, \end{array}\right.
\nonumber
\end{align}
\end{widetext}
where $|q_j|=\sqrt{|q_0^2-2\me\omega_{j0}/\hbar|}$, $\theta_{jl}=1$ if either $p_l-h/2<|q_j|<p_l+h/2$ or $p_l-h/2<-|q_j|<p_l+h/2$, and $\theta_{jl}=0$ otherwise. In addition, we use the explicit expressions offered in Sec.\ \ref{small} for the coupling coefficients $G_{qj,q'j'}$ associated with selected excitations of well-defined multipolar symmetries. Incidentally, $\Delta_{j,ll'}$ may diverge in the unlikely event that $q_{j\neq0}$ coincides with $p_l\pm h/2$ for any $l$, a problem that we avoid by slightly changing $N$.

We apply this procedure to study two different types of samples in the main text, for which the evaluation of the coupling matrices $G_{jj'}$ and the numerical solution of Eq.\ (\ref{Mjblocks}) are simplified by the following considerations:
\begin{itemize}
\item {\it Two-level sample.} For a system comprising two states $j=0$ and 1, we have $G_{10}=G_{01}^\dagger$ and $G_{00}=G_{11}=0$. The solution of Eq.\ (\ref{Mjblocks}) for the excited state vector is $M_1=\big(1-S_{10}\cdot S_{01}\big)^{-1}\!\!\cdot g_1$ and requires performing one matrix inversion and one product between square matrices, each of them involving $\sim N^3$ complex-number multiplications.
\item {\it Single-boson sample.} Coupling to a bosonic mode of frequency $\omega_b$ can also be described using the present formalism. The bosonic Fock states $|j\rangle$ appear at equally spaced frequencies given by $\omega_{j0}=j\omega_b$, whereas the coupling matrix elements only connect consecutive states, and consequently, we have
\begin{align}
G_{jj'}=\left\{\begin{array}{rl}
0,                          &\quad\quad |j-j'|\neq1, \\
\sqrt{j}\,G_{10},           &\quad\quad j'=j-1,      \\
\sqrt{j+1}\,G_{10}^\dagger, &\quad\quad j'=j+1.
\end{array}\right.
\label{Gboson}
\end{align}
These matrices are expressed in terms of $G_{10}$, which describes $0\rightarrow1$ transitions. Then, Eq.\ (\ref{Mjblocks}) becomes a block-tridiagonal system, which we solve following the standard forward-sweep, backward-substitution method, requiring a total of $\sim (n+1)\times N^3$ complex-number multiplications when the boson ladder is cut at $j=n$.
\end{itemize}
Incidentally, for these two types of samples, assuming the system to be initially prepared in the ground state $j=0$ and considering a given final state $j=j_1$, all of the terms in the scattering series obtained from Eq.\ (\ref{Mjblocks}) by Taylor expanding the self-consistent equation $M=(1+S)^{-1}\cdot g=g-S\cdot g + S^2\cdot g+\cdots$ present the same difference between the net number of up ($j\rightarrow j+1$) and down ($j\rightarrow j-1$) jumps, so the overall probabilities are left unchanged when $G_{10}$ is multiplied by an arbitrary phase factor $\ee^{\ii\varphi}$ (i.e., $G_{jj'}$ introduces a factor $\ee^{-\ii\varphi}$ or $\ee^{\ii\varphi}$ for up or down jumps, respectively, and the final amplitude is modified by an overall amplitude $\ee^{-\ii j_1\varphi}$).

Using the methods discussed above, we find convergent results within the scale of the plots in the main text using $2N+1\sim500$ discretization points and a wave vector range $\Delta$ of a few times $q_0\big(1-\sqrt{1-\omega_{10}/\epsilon_{q_0}}\big)$. In addition, for sufficiently large electron energy $\varepsilon_{q_0}\gg\omega_{10}$, this procedure produces results in excellent quantitative agreement with the solution found in the nonrecoil approximation [see Eq.\ (\ref{fgfe}) in Sec.\ \ref{nonrecoilapprox} below].

\section{Nonrecoil approximation}
\label{nonrecoilapprox}

A simpler solution is obtained if the energy spread of the incident e-beam and the sample excitation energies are both small compared with the average electron energy. We can then approximate $\varepsilon_q-\varepsilon_{q'}\approx(q-q')v$, where $v=\sum_qv_q\big|\alpha_q^0\big|^2$ is the average electron velocity. In addition, we use the fact that the coefficients $G_{qj,q'j'}$ only depend on the difference of wave vectors $q-q'$ \cite{noteonGqq}. Applying these considerations to Eq.\ (\ref{psi}), we can write
\begin{align}
\langle z|\Psi(t)\rangle=\frac{\ee^{\ii\bar{q}z-\ii\varepsilon_{\bar{q}}t}}{\sqrt{2\pi}}\sum_j\varphi_j(z,t)\,\ee^{-\ii\omega_jz/v}\,|j\rangle,
\nonumber
\end{align}
where $\bar{q}$ is the average wave vector, and we define the electron wave function
\begin{align}
\varphi_j(z,t)&=\int\!dq\;\ee^{\ii (q-\bar{q}+\omega_j/v)(z-vt)}\alpha_{qj}(t) \nonumber
\end{align}
associated with the each sample state $j$. Likewise, multiplying both sides of Eq.\ (\ref{dadt}) by $\ee^{\ii (q-\bar{q}+\omega_j/v)(z-vt)}$ and integrating over $q$, we obtain
\begin{align}
(\partial_t+v\partial_z)\,\varphi_j(z,t)&=-\ii\sum_{j'}G_{jj'}(z)\,\ee^{\ii\omega_{jj'}z/v}\,\varphi_{j'}(z,t), \nonumber
\end{align}
where the coupling coefficient
\begin{align}
G_{jj'}(z)=\int\!dq\;G_{qj,0j'}\,\ee^{\ii qz}
\label{Gjjz}
\end{align}
is expressed in the real-space representation.

These equations admit a solution of the form $\varphi_j(z,t)=\phi^0(z-vt)\,\ee^{\ii\omega_0(z/v-t)}\,f_j(z)$, consisting of an overall factor than only depends on $z-vt$, accompanied by space-dependent functions $f_j(z)$ that satify
\begin{align}
\frac{d\,f_j(z)}{dz}&=-\frac{\ii}{v}\sum_{j'} G_{jj'}(z)\;\ee^{\ii\omega_{jj'}z/v}\,f_{j'}(z).  \label{fgfe}
\end{align}
Incidentally, we have explicitly indicated a factor $\ee^{\ii\omega_0(z/v-t)}$ in $\varphi_j(z,t)$ to simplify the initial condition for a sample prepared in the $|0\rangle$ state (see below). Also, we can set $\phi^0(z-vt)=\int\!dq\; \alpha_q^0\;\ee^{\ii (q-\bar{q})(z-vt)}$, as suggested by the incident electron wave function in Eq.\ (\ref{psi0}), which takes the form $\psi^0(z,t)=\big(\ee^{\ii\bar{q}z-\ii\varepsilon_{\bar{q}} t}/\sqrt{2\pi}\big)\phi^0(z-vt)$ in the nonrecoil approximation. Therefore, the combined electron-sample wave function reduces to
\begin{align}
\langle z|\Psi(t)\rangle=\psi^0(z,t) \sum_j f_j(z)\,\ee^{-\ii\omega_{j0}(z/v-t)}\,\ee^{-\ii\omega_jt}\,|j\rangle,
\label{zpsipsi0}
\end{align}
which is a superposition of time-dependent states $\ee^{-\ii\omega_jt}\,|j\rangle$, each of them accompanied by an electron wave function $\psi^0(z,t)\,\ee^{-\ii\omega_{j0}(z/v-t)}$ that reflects the associated change in electron momentum by $-\hbar\omega_{j0}/v$. Once $f_j(z)$ is obtained by solving Eq.\ (\ref{fgfe}) with the initial conditions $f_j(-\infty)=\delta_{j0}$ (i.e., the sample in the $|0\rangle$ state), the post-interaction probabilities are finally given by $P_j=\big|f_j(\infty)\big|^2$.

This analysis can be readily applied to any initial pure state $|j_i\rangle$ by simply substituting $\omega_0$ by $\omega_{j_i}$ in these expressions, and also extended to start with a coherent superposition of sample states $\sum_{j_i}a_{j_i}\ee^{-\ii\omega_{j_i}t}|j_i\rangle$ by separately propagating each initial state $|j_i\rangle$ and then weighting the resulting $f_j(z)$ with the coefficients $a_{j_i}$, so that we obtain the wave function
\begin{align}
\langle z|\Psi(t)\rangle=\psi^0(z,t) \sum_{jj_i} F_{jj_i}(z)\,a_{j_i}\,\ee^{-\ii\omega_{jj_i}(z/v-t)}\,\ee^{-\ii\omega_jt}\,|j\rangle,
\nonumber
\end{align}
where $F_{jj_i}(z)$ is given by the coefficient $f_j(z)$ obtained from Eq.\ (\ref{fgfe}) with the initial conditions $f_j(-\infty)=\delta_{jj_i}$. For incident nonochromatic electrons, the probability of finding the sample in state $j$ after the interaction has taken place becomes $P_j=\sum_{j_i}\big|F_{jj_i}(\infty)\,a_{j_i}\big|^2$, whereas the electron energy-loss probability reduces to $\Gamma_{\rm EELS}(\omega)=\sum_{jj_i}\big|F_{jj_i}(\infty)\,a_{j_i}\big|^2\delta(\omega-\omega_{jj_i})$, which in practice needs to be convoluted with the zero-loss peak of the microscope.

Again, we corroborate the trivial role played by the incident electron wave function: in the nonrecoil approximation, the final total wave function is the incident electron wave function multiplied by factor [the $j$ sum in Eq.\ (\ref{zpsipsi0})] that only depends on the incident electron state through the velocity $v$. From a quantum-optics perspective \cite{paper339}, this result reflects the fact that the electron acts on the sample as a classical source [i.e., regarding Eq.\ (\ref{fgfe}) as the optical Bloch equations of the system] if its velocity is taken to be constant (i.e., in the nonrecoil approximation), so that starting  with $j=0$, and in virtue of energy conservation, the interaction simply causes the electron wave function to undergo rigid shifts $\approx-\hbar\omega_{j0}/v$ in momentum when the sample is excited to a state $j\neq0$.

For reference, the linear probability for $j\neq0$ now corresponds to the solution of Eq.\ (\ref{fgfe}) obtained by replacing $f_{j'}(z)$ by $\delta_{j'0}$ in the right-hand side, and consequently, we find $f_j(\infty)=(-\ii/v)\int dz\, G_{j0}(z)\;\ee^{\ii\omega_{j0}z/v}=(-2\pi\ii/v)\,G_{q_jj,q_00}$ with $q_j=q_0-\omega_{j0}/v$ [i.e., the inverse Fourier transform of Eq.\ (\ref{Gjjz})], which leads to $P_j^{\rm lin}=(4\pi^2/v^2)\big|G_{q_jj,q_00}\big|^2=(4\pi^2/v^2)\big|G_{0j,\omega_{j0}/v,0}\big|^2$. This result agrees with Eq.\ (\ref{P1lin}) when neglecting the backscattering term $\propto\big|G_{-q_jj,q_00}\big|^2$ and setting $v_q\approx v$.

\subsection{Interaction with a bosonic mode}

The solution to Eq.\ (\ref{fgfe}) becomes analytical for a sample hosting a single boson mode of frequency $\omega_b$, which comprises an infinite number of discrete Fock states $|j\rangle$ of frequencies $j\omega_b$ with $j=0,1,\dots$, coupled by coefficients $G_{jj'}(z)$ that satisfy Eq.\ (\ref{Gboson}). Indeed, for such bosonic system, Eq.\ (\ref{fgfe}) reduces to
\begin{align}
\frac{d\,f_j(z)}{dz}&=\sqrt{j}\,u^*(z)\,f_{j-1}(z)-\sqrt{j+1}\,u(z)\,f_{j+1}(z) \nonumber
\end{align}
with $u(z)=(\ii/v)G_{10}^*(z)\ee^{-\ii\omega_bz/v}$, so it admits the closed-form solution \cite{paper339} $f_j(z)=\ee^{\ii\chi(z)}\ee^{-|\beta_0(z)|^2/2}\big[\beta_0^*(z)\big]^j/\sqrt{j!}$, where $\chi(z)=\int_{-\infty}^z dz'\int_{-\infty}^{z'}dz''\,{\rm Im}\{u^*(z')u(z'')\}$ is a global phase and $\beta_0(z)=\int_{-\infty}^zdz'\,u(z')$. The boson is then evolving as a coherent state of varying amplitude $\beta_0(z)$, featuring a Possonian distribution of occupation numbers $P_j(z)=|f_j(z)|^2=\ee^{-|\beta_0(z)|^2}\big|\beta_0(z)\big|^{2j}/j!$ and average population $\sum_{j=1}^\infty j\,P_j(z)=|\beta_0(z)|^2$ that leads to the post-interaction value
\begin{align}
\langle j\rangle&=\sum_{j=0}^\infty j\,P_j=|\beta_0(\infty)|^2 \nonumber\\
&=\frac{1}{v^2}\left|\int dz\,G_{10}(z)\ee^{\ii\omega_bz/v}\right|^2 \nonumber\\
&=\frac{4\pi^2}{v^2}\big|G_{01,\omega_b/v,0}\big|^2=P_1^{\rm lin},
\nonumber
\end{align}
where the integral is again identified with the inverse Fourier transform connecting $G_{10}(z)$ to $G_{q1,00}=G_{01,-q,0}$ [see Eq.\ (\ref{Gjjz})].

\section{Nonretarded coupling coefficients for small samples}
\label{small}

In the nonretarded limit, the electron-sample interaction is mediated by the Coulomb potential, so the corresponding matrix elements in Eq.\ (\ref{H}) read
\begin{align}
G_{qj,q'j'}&=\frac{1}{\hbar}\langle qj|\hat{\mathcal{H}}|q'j'\rangle \nonumber\\
&=-\frac{e}{2\pi\hbar}\int dz_e\; \ee^{\ii(q'-q)z_e} \int d^3\rb \frac{\langle j|\hat{\rho}(\rb)|j'\rangle}{|\rb_e-\rb|} \nonumber\\
&=-\frac{e}{\pi\hbar}\int d^3\rb\;\langle j|\hat{\rho}(\rb)|j'\rangle \label{gqq}\\
&\quad\quad\quad\times K_0(|q-q'||\Rb_e-\Rb|)\;\ee^{\ii(q'-q)z},
\nonumber
\end{align}
where $\hat{\rho}(\rb)=\sum_i q_i \,\delta(\rb-\rb_i)$ is the charge density operator, expressed as a sum over electrons and ions of charges $q_i$ and positions $\rb_i$ in the specimen, and $\Rb_e$ is the e-beam impact parameter, set to a fixed value in accordance with assumption {\it \red {\rm(}i\rm{)}} in Sec.\ \ref{sec1}. As anticipated in Sec.\ \ref{nonrecoilapprox}, $G_{qj,q'j'}$ only depends on the wave vector difference $q-q'$. Incidentally, we neglect coupling terms with $j=j'$, although one could conceivably imagine a system in which the time-dependent energy shifts produced by the presence of the electron in the states $j$ could affect the excitation dynamics, as recently investigated for the nonlinear interaction between low-energy free electrons and nanographenes \cite{paper350}.

For a small sample compared with the impact parameter $R_e$, the coupling coefficients in Eq.\ (\ref{gqq}) can be approximated by only retaining the first nonvanishing contribution to the Taylor expansion of $K_0(|q-q'||\Rb_e-\Rb|)\ee^{\ii(q'-q)z}$ around $\rb=0$. The zeroth-order term cancels for a neutral sample (i.e, $\int d^3\rb\;\hat{\rho}(\rb)=\sum_iq_i=0)$, while the linear term yields
\begin{align}
G_{qj,q'j'}=-\frac{e}{\pi\hbar}\,\pb_{jj'}\cdot&\bigg[|q-q'|K_1(|q-q'|R_e)\,\hat\Rb_e \label{Gdipole}\\
&+\ii(q'-q)\,K_0(|q-q'|R_e)\,\zz\bigg],
\nonumber
\end{align}
where $\pb_{jj'}=\int d^3\rb\;\langle j|\hat{\rho}(\rb)|j'\rangle\;\rb$ is the transition dipole. Inserting this result in Eq.\ (\ref{Gjjz}) to obtain the space-dependent interaction in the nonrecoil approximation, we find the expected dipole potential $G_{jj'}(z)=-(e/\hbar)\,\pb_{jj'}\cdot(\Rb_e+z\,\zz)/(R_e^2+z^2)^{3/2}$.

A more general treatment that is suitable for dealing with multipolar excitations can be followed by expanding the Coulomb potential in Eq.\ (\ref{gqq}) using spherical harmonics as \cite{J99} \[\frac{1}{|\rb_e-\rb|}=\sum_{l=0}^\infty\sum_{m=-l}^l \frac{4\pi}{2l+1}\,(r^l/r_e^{l+1})\,Y_{lm}(\rr_e)Y_{lm}^*(\rr)\] under the assumption that $r_e>r$ (i.e., provided the sample can be inscribed in a sphere that is not intersected by the e-beam). Inserting this expansion into Eq.\ (\ref{gqq}) and using the analytical expression for the $z_e$ integral of $\ee^{\ii(q'-q)z_e}Y_{lm}(\rr_e)/r_e^{l+1}$ derived in Ref.\ \cite{paper021}, we obtain
\begin{widetext}
\begin{align}
G_{qj,q'j'}=-\frac{e}{\pi\hbar}\sum_{l=1}^\infty\sum_{m=-l}^l Q_{lm,jj'}
\frac{(-\ii)^{l+m}\;\ee^{\ii m\varphi_{\Rb_e}}}{\sqrt{(l-m)!(l+m)!}}(q-q')^lK_m(|q-q'|R_e)
\times\left\{\begin{matrix} 1, &\quad q-q'>0, \\ (-1)^{m}, &\quad q-q'<0, \end{matrix}\right. \label{multipolar}
\end{align}
\end{widetext}
where
\begin{align}
Q_{lm,jj'}=\sqrt{\frac{4\pi}{2l+1}}\int d^3\rb \;\langle j|\hat{\rho}(\rb)|j'\rangle \,r^l Y_{lm}^*(\rr),
\nonumber
\end{align}
$\varphi_{\Rb_e}$ is the azimuthal angle of $\Rb_e$, and we eliminate the $l=0$ term because it vanishes due to charge neutrality in the sample. Incidentally, we follow the notation of Ref.\ \cite{M1966} for the spherical harmonics $Y_{lm}(\rr)$, which differs by a factor $(-1)^m$ from that in Ref.\ \cite{AS1972}. Detailed inspection shows that the $l=1$ term in this expression reproduces the result in Eq.\ (\ref{Gdipole}), with $Q_{1m,jj'}$ reducing to the components of the transition dipole.

\begin{table*}[t]
\begin{tabular}{c|c|c|r|l} \hline
$\begin{matrix}\text{multipolar}\\ \text{order}\end{matrix}$                                    \quad&\quad
$\begin{matrix}\text{charge density}\\ \text{symmetry}\end{matrix}$                             \quad&\quad
$\begin{matrix}\text{charge density}\\ \langle j|\hat{\rho}(\rb)|j'\rangle\propto \end{matrix}$ \quad&\quad
$g_{jj'}\times G_{jj'}(z)$                                                                   \quad\quad\quad\quad\quad&\quad
$g_{jj'}\times G_{qj,0j'}$ \\
\hline
&&&&\\dipole    \quad&\quad $p_x$         \quad&\quad $x/r$           \quad&\quad $\pi\,R_e/(R_e^2+z^2)^{3/2}$ \quad\quad&\quad $|q|\;K_1\big(|q|R_e\big)$ \\
&&&&\\dipole    \quad&\quad $p_z$         \quad&\quad $z/r$           \quad&\quad $\ii\pi\,z/(R_e^2+z^2)^{3/2}$ \quad\quad&\quad $q\;K_0\big(|q|R_e\big)$ \\
&&&&\\quadrupole\quad&\quad $d_{z^2}$     \quad&\quad $3z^2/r^2$      \quad&\quad $\pi(R_e^2-2z^2)/(R_e^2+z^2)^{5/2}$ \quad\quad&\quad $q^2\;K_0\big(|q|R_e\big)$ \\
&&&&\\quadrupole\quad&\quad $d_{xz}$      \quad&\quad $xz/r^2$        \quad&\quad $3\ii\pi\,z\,R_e/(R_e^2+z^2)^{5/2}$ \quad\quad&\quad $q^2\;K_1\big(|q|R_e\big)\;{\rm sign}\{q\}$ \\
&&&&\\quadrupole\quad&\quad $d_{x^2-y^2}$ \quad&\quad $(x^2-y^2)/r^2$ \quad&\quad $3\pi\,R_e^2/(R_e^2+z^2)^{5/2}$ \quad\quad&\quad $q^2\;K_2\big(|q|R_e\big)$ \\
&&&&\\ \hline
\end{tabular}
\caption{Coupling coefficients for dipolar and quadrupolar $j'\rightarrow j$ transitions with different symmetries. The e-beam moves along the $z$ axis and intersects the $x$ axis at a distance $R_e$ from the sample. A $q$- and $z$-independent normalization constant $g_{jj'}$ is introduced in the two rightmost columns, encapsulating the dependence on the radial matrix elements. We note that $G_{jj'}(z)$ is obtained from $G_{qj,0j'}(z)$ by using Eq.\ (\ref{Gjjz}). The two remaining quadrupolar transitions $d_{xy}$ and $d_{yz}$ do not couple to the electron with the specified trajectory.}
\label{TableS1}
\end{table*}

Beyond dipoles, we also explore quadrupolar excitations characterized by charge densities $\langle j|\hat{\rho}(\rb)|j'\rangle$ with a spatial angular dependence given by $Y_{20}\propto(3z^2/r^2-1)$, $(Y_{2,-1}-Y_{21})\propto xz/r^2$, and $(Y_{2,-2}+Y_{22})\propto(x^2-y^2)/r^2$, as found, for example, in transitions from a hydrogenic $s$ orbital to $d_{z^2}$, $d_{xz}$, and $d_{x^2-y^2}$ states. We take $\varphi_{\Rb_e}=0$ (i.e., the e-beam crosses the $x$ axis), so that the other two possible quadrupolar excitations ($d_{xy}$ and $d_{yz}$) do not couple to the electron because of symmetry mismatch. In Table\ \ref{TableS1}, we summarize the associated coupling coefficients in both momentum and real-space representations [$G_{qj,0j'}$ and $G_{jj'}(z)$, respectively] for the nonzero dipolar and quadrupolar transitions under the noted conditions. In this work, we use $G_{qj,0j'}$ in combination with Eq.\ (\ref{Mjblocks}) to produce numerical nonlinear results including recoil effects, while numerical integration of Eq.\ (\ref{fgfe}) with $G_{jj'}(z)$ as input allows us to compute excitation probabilities in the nonrecoil approximation. We remark that the normalization constants $g_{jj'}$ in Table\ \ref{TableS1} encapsulate all the factors that accompany the $(q-q')^lK_m(|q-q'|R_e)$ dependence for each angular symmetry in Eq.\ (\ref{multipolar}), and in particular, they are proportional to the multipolar transition strength $Q_{lm,jj'}$.

In the main text, rather than specifying $g_{jj'}$, we express our results as a function of the first-order (linear) excitation probability $P_1^{\rm lin}$ given by Eq.\ (\ref{P1lin}), so we take
\begin{align}
|g_{jj'}|=2\pi\sqrt{\frac{|g_{jj'}G_{q_jj,q_00}|^2+|g_{jj'}G_{-q_jj,q_00}|^2}{v_{q_0}v_{q_j}P_j^{\rm lin}}}
\label{gjj}
\end{align}
with $g_{jj'}G_{\pm q_jj,q_00}\equiv g_{jj'}G_{\pm q_j-q_0,j,00}$ explicitly given by the rightmost column of Table\ \ref{TableS1} with $q=\pm q_j-q_0$. This prescription needs to be modified in the nonrecoil approximation, as we argue in Sec.\ \ref{nonrecoilapprox}, so we calculate $|g_{jj'}|$ after eliminating the backscattering term $|g_{jj'}G_{-q_jj,q_00}|^2$ in Eq.\ (\ref{gjj}) to write $|g_{jj'}|=(2\pi/v)|g_{jj'}G_{q_jj,q_00}|\big/\big(P_j^{\rm lin}\big)^{1/2}$.

\bibliographystyle{apsrev} 
\bibliography{../../../bibtex/refsL.bib} 

\begin{thebibliography}{48}
\bibitem[{not({\natexlab{a}})}]{noteonGqqbis}
\bibinfo{note}{This result comes from the $\delta$-function contribution of
  $(\varepsilon_{q'q_0}+\omega_{j'0}-\ii0^+)^{-1}={\rm
  PV}\left\{(\varepsilon_{q'q_0}+\omega_{j'0})^{-1}\right\}+\ii\pi\delta(\varepsilon_{q'q_0}+\omega_{j'0})$
  to Eq.\ (\ref{Njfinal}), where PV stands for the principal value, which
  vanishes identically.}
\end{thebibliography}

\end{document}